\documentclass[final,twocolumn]{elsarticle}

\usepackage{amssymb}
\usepackage{amsmath}
\usepackage[capitalise]{cleveref}
\usepackage{graphicx}
\usepackage{caption}
\usepackage{subcaption}

\DeclareMathOperator{\erf}{erf}
\newcommand{\unit}[1]{\,\textrm{#1}}

\journal{Continental Shelf Research}

\begin{document}

\begin{frontmatter}

\title{Spectral Methods for Coastal-Trapped Waves and Instabilities in a Background Flow}

\author[inst1]{Matthew N. Crowe\corref{cor1}}
\author[inst2]{Edward R. Johnson}

\cortext[cor1]{Matthew.Crowe2@newcastle.ac.uk}

\affiliation[inst1]{organization={School of Mathematics, Statistics and Physics},
            addressline={Newcastle University}, 
            city={Newcastle upon Tyne},
            postcode={NE1 7RU}, 
            country={UK}}

\affiliation[inst2]{organization={Department of Mathematics},
            addressline={University College London}, 
            city={London},
            postcode={WC1E 6BT}, 
            country={UK}}

\begin{abstract}
Here we present a numerical method for finding non-hydrostatic coastal-trapped wave and instability solutions to the non-hydrostatic Boussinesq equations in the presence of a background flow and complicated coastal topography. We use spectral methods to discretise the two-dimensional eigenvalue problem and solve the resulting discrete problem by standard methods. Our approach is applied to three examples and shown to be consistent with previous numerical and analytical results. In particular, we show that our method is able to reliably identify coastal-trapped wave solutions that correspond to waves seen in realistic simulations of the Southeast Greenland shelf.
\end{abstract}


\end{frontmatter}


\section{Introduction}
\label{sec:intro}

The term `coastal-trapped waves' (CTWs) describes those oceanographic waves which exist in coastal regions and propagate in an alongshore direction \cite{Mysak_1980}. They exist primarily due to a balance between the Coriolis force and the normal reaction force from a coastal boundary and, as such, their direction of propagation is determined by the sign of the Coriolis parameter; waves in the Northern (Southern) hemisphere propagate with the coastline to their right (left). CTWs play a major role in coastal dynamics and influence a wide range of phenomena. For example, CTWs travel quickly and hence provide a mechanism for transferring information and energy rapidly around an ocean basin and governing coastal sea levels \cite{Hughes_et_al_2019}. Similarly, they are important for vorticity generation and dissipating ocean energy along coastlines \cite{DEWARHOGG,DEREMBLEETAL}. CTWs are commonly generated by wind \cite{GillSchumann74} or tides \cite{LeBlondM78} and energy transfer between different types of CTW can occur due to interactions with topography \cite{Weber_Isachsen_2023}. Another mechanism for CTW generation is the interaction of flow features, such as vortices, with coastal topography \cite{Crowe_Johnson_2021,Crowe_Johnson_23}.

Two important types of CTW are Kelvin waves \cite{Thomson_1880} and shelf waves \cite{LeBlondM78}. Kelvin waves are a type of non-dispersive, rotating, gravity (or internal gravity) wave which propagate along vertical boundaries and are restored by the effects of gravity on deformed density surfaces. Conversely, shelf waves are a form of topographic Rossby wave which require a sloping boundary and are restored by the conservation of potential vorticity. As such, they are dispersive and can be observed in models with constant density and no free surface. In the ocean, Kelvin waves and shelf waves exist simultaneously due to the co-existence of complicated topographic features and a density stratification \cite{Huthnance_78}. CTWs are further modified by the existence of a background flows, such as commonly observed coastal jets \cite{Gelderloos_et_al_21}.

A concept closely related to a wave is that of an instability; where a disturbance grows with time rather than propagating. Instabilities commonly arise due to wave interactions in the presence of a geostrophically balanced flow, for example the classical mechanism for barotropic instability \cite{Kuo_1949} is the interaction between counter-propagating Rossby waves. Baroclinic instability \cite{EADY} may similarly be described through the interaction of edge waves. Instabilities arising due to the destabilisation of CTWs by a background flow may be important for the dynamics of flows over coastal topography \cite{Mysak_et_al_81,Brink_2012} and are believed to be related to an increase in eddy activity over continental shelves \cite{Brink_2016}.

Typically, wave and instability solutions are identified by assuming sinusoidal behaviour in space and time and solving an algebraic problem for the frequency or growth-rate in terms of the spatial wavenumber and problem parameters. This procedure is complicated by spatial inhomogeneities which prevent the use of sinusoidal dependence in that direction. The CTW problem is especially difficult as sloping topography introduces inhomogeneity in both the vertical and offshore directions, resulting in an eigenvalue problem for the frequency in which the eigenfunctions depend on two spatial coordinates. This problem is not, in general, analytically tractable and has been approached numerically by various authors. Huthnance \cite{Huthnance_78} used spatial discretisation via a finite difference method and inverse-iteration to solve this problem. Johnson \& Rodney \cite{Johnson_Rodney_2011} later used exponentially accurate spectral methods to discretise the system, resulting in a much smaller numerical problem which allowed CTW modes to be easily identified without needing an accurate initial guess for the frequency. Another commonly used approach is that of Brink \cite{Brink_2006} which uses simplex minimisation to solve the problem of Huthnance \cite{Huthnance_78} with the inclusion of bottom friction and background flow.

In this paper, we extend the method of \cite{Johnson_Rodney_2011} to include a background flow, non-hydrostatic effects, and an increased versatility in the offshore discretisation. We begin in \cref{sec:setup} by formulating the CTW problem and describing the assumptions inherent in our model. In \cref{sec:num_meth} we outline the numerical approach used to solve this system and some techniques for identifying the correct solutions in the case of a complicated topography or background flow. To demonstrate the effectiveness of our approach, we present three examples in \cref{sec:examples}, covering cases of idealised waves, idealised instabilities and CTWs over realistic topography. In particular, we consider the case of CTWs on the Southeast Greenland shelf, motivated by the recent work of Gelderloos et. al. \cite{Gelderloos_et_al_21}. Finally, in \cref{sec:diss_conc} we summarise and discuss our method and results.

\section{Problem Setup}
\label{sec:setup}

Consider a three-dimensional region of fluid bounded by a vertical wall at $y = 0$ and a bottom boundary at $z = -H(y)$. The fluid is rotating with Coriolis parameter $f$ and vertically stratified with Buoyancy frequency $N = N(z)$. An along-shelf flow background flow is included with velocity $U(y,z)$ in the $x$ direction. The linearised Boussinesq equations are
\begin{align}
\label{eq:u_eqn}
\left(\partial_t + U\partial_x\right) u + U_y v + U_z w - fv + \partial_x p = 0,\\
\left(\partial_t + U\partial_x\right)  v + fu + \partial_y p = 0,\\
\delta_h\left(\partial_t + U\partial_x\right) w - b + \partial_z p = 0,\\
\label{eq:b_eqn}
\left(\partial_t + U\partial_x\right) b + N^2 w = 0,\\
\label{eq:m_eqn}
\partial_x u + \partial_y v + \partial_z w = 0.
\end{align}
Here $\delta_h$ is the hydrostatic parameter and may be set to $0$ to consider hydrostatic solutions or $1$ to include non-hydrostatic effects. We impose no-flow through the bottom boundary
\begin{equation}
w + H_y v = 0 \quad \textrm{on} \quad z = -H(y),
\end{equation}
and no-flow through the vertical wall
\begin{equation}
v = 0 \quad \textrm{on} \quad y = 0.
\end{equation}
Our final boundary condition is obtained from the free surface condition
\begin{equation}
w = \left(\partial_t+U\partial_x\right) \eta,
\end{equation}
where $\eta = p/g$ is the (small) free surface height and $g$ is the gravitational acceleration. This may be combined with \cref{eq:b_eqn} to give the linearised surface condition
\begin{equation}
b + \delta_a \frac{N^2}{g} p = 0 \quad \textrm{on} \quad z = 0,
\end{equation}
where $\delta_a$ in the free-surface parameter. Setting $\delta_a = 0$ allows us to study the rigid lid case whereas $\delta_a = 1$ corresponds to a free surface and supports surface waves. Far from the shelf region, we expect wave-like perturbations to decay so we impose the far field condition
\begin{equation}
(u,v,w,b,p) \to 0 \quad \textrm{as} \quad y \to \infty.
\end{equation}

To proceed, we consider normal mode solutions in the along-shelf ($x$) direction of the form
\begin{equation}
\phi = \hat{\phi}(y,z) \exp[i(kx-\omega t)],
\end{equation}
so \cref{eq:u_eqn}-\cref{eq:m_eqn} can be written in the form
\begin{equation}
\label{eq:EVP}
\omega\boldsymbol{D}\hat{\boldsymbol{\phi}} = \boldsymbol{L} \hat{\boldsymbol{\phi}},
\end{equation}
where
\begin{equation}
\hat{\boldsymbol{\phi}} = \begin{pmatrix}
\hat{u} \\ -i\hat{v} \\ -i\hat{w} \\ \hat{b} \\ \hat{p}
\end{pmatrix},
\end{equation}
\begin{equation}
\boldsymbol{D} = \begin{pmatrix}
1 & 0 & 0 & 0 & 0 \\
0 & -1 & 0 & 0 & 0 \\
0 & 0 & -\delta_h & 0 & 0 \\
0 & 0 & 0 & 1 & 0 \\
0 & 0 & 0 & 0 & 0
\end{pmatrix},
\end{equation}
and
\begin{equation}
\boldsymbol{L} = \begin{pmatrix}
kU & U_y-f & U_z & 0 & k \\
f & -kU & 0 & 0 & \partial_y \\
0 & 0 & -\delta_h kU & -1 & \partial_z \\
0 & 0 & N^2 & kU & 0 \\
k & \partial_y & \partial_z & 0 & 0
\end{pmatrix}.
\end{equation}
Solving for $-i\hat{v}$ and $-i\hat{w}$ rather than $\hat{v}$ and $\hat{w}$ ensures that the operators $\boldsymbol{D}$ and $\boldsymbol{L}$ are real. This system is solved subject to the boundary conditions
\begin{align}
\hat{w} + H_y\hat{v} = 0 & \quad \textrm{on} \quad z = -H(y),\\
\hat{v} = 0 & \quad \textrm{on} \quad y = 0,\\
\hat{b} + \delta_a \frac{N^2}{g} \hat{p} = 0 & \quad \textrm{on} \quad z = 0,\\
(\hat{u},\hat{v},\hat{w},\hat{b},\hat{p}) \to 0 & \quad \textrm{as} \quad y \to \infty.
\end{align}

\section{Numerical Method}
\label{sec:num_meth}

We now solve \cref{eq:EVP} numerically using a spectral collocation approach similar to that of \cite{Johnson_Rodney_2011}. We opt to solve \cref{eq:EVP} directly rather than converting it to a single equation for $\hat{p}$ as the inclusion of the non-hydrostatic terms in the vertical momentum equation means that such a conversion does not reduce the size of the resulting numerical problem.

\subsection{Coordinate transformation}

In order to use spectral collocation methods, we first transform to a rectilinear grid using the coordinate transformation
\begin{equation}
\label{eq:def_transf}
(\lambda,\zeta) = \left(y,\frac{z}{H(y)}\right),
\end{equation}
where $\lambda \in [0,\infty)$ and $\zeta \in [-1, 0]$. The derivatives in $(y,z)$ space transform as
\begin{equation}
\label{eq:deriv_transf}
\partial_y = \partial_\lambda - \frac{\zeta H'(\lambda)}{H(\lambda)}\partial_\zeta, \quad \partial_z = \frac{1}{H(\lambda)} \partial_\zeta.
\end{equation}
The operators $\partial_y$ and $\partial_z$ are calculated in discretised $(\lambda,\zeta)$ space rather than applying the coordinate transform in \cref{eq:def_transf} to 
$\boldsymbol{D}$ and $\boldsymbol{L}$ directly.

\subsection{The numerical grid}

We now discretise our spatial coordinates $\lambda$ and $\zeta$ using $N_\lambda$ points in the offshore direction and $N_\zeta$ points in the vertical. We use the spectral collocation method which works by approximating a function as a sum over a (finite) set of polynomial basis functions. The points used correspond to the roots of the highest order polynomial in the basis. Differentiation may then be performed exactly on each basis polynomial by writing its derivative as a sum over the full set. Therefore, differentiation corresponds to multiplying the vector of function values by a spectral collocation matrix. To ensure fast convergence and avoid spurious oscillations, sensible choices of points have an uneven distribution with a higher density of points near the boundaries \cite{Trefethen}.

Chebyshev points are a common choice for finite domains and are used here for discretising $\zeta$ and the corresponding derivative. In the $\lambda$ direction, we need an approach which is suitable for a semi-infinite domain. Laguerre points are the natural choice here as they correspond to an expansion in terms of Laguerre polynomials scaled by a decaying exponential. However, since Laguerre points are clustered near $\lambda = 0$, an offshore jet---or complicated bottom topography far from the wall---will not be well resolved unless a high number of points in used. Instead, we use a composite grid consisting of several segments of Chebyshev points joined to a single segment of Laguerre points for $\lambda$. Derivatives are calculated using spectral collocation in each segment and at the intersection of neighbouring segment, the derivative is taken to be the average of the derivatives on the left and right segments. More advanced methods may be used to join two segments; however, we find that our approach resolves derivatives at intersections with a similar error to that of the rest of the domain.

\subsection{Solving the discretised problem}

We may now discretise \cref{eq:EVP} using our discrete domain and derivatives in $(\lambda,\zeta)$ space giving the system
\begin{equation}
\omega \tilde{\boldsymbol{D}} \tilde{\boldsymbol{\phi}} = \tilde{\boldsymbol{L}} \tilde{\boldsymbol{\phi}},
\end{equation}
where $\tilde{\boldsymbol{D}}$ and $\tilde{\boldsymbol{L}}$ are matrices of size $(5N_\lambda N_\zeta)^2$ and $\tilde{\boldsymbol{\phi}}$ is a state vector of length $5N_\lambda N_\zeta$. This problem may be solved using a standard generalised eigenvalue solver such as the Matlab `eigs' function. In order to identity solutions close to some eigenvalue, $\omega_0$, we write $\omega = \omega_0 + \omega'$ and solve the modified system
\begin{equation}
\tilde{\boldsymbol{D}} \tilde{\boldsymbol{\phi}} = \frac{1}{\omega'}[\tilde{\boldsymbol{L}} -\omega_0 \tilde{\boldsymbol{D}}] \tilde{\boldsymbol{\phi}}
\end{equation}
for eigenvalue $1/\omega'$. Searching for eigenvalues, $1/\omega'$, with the largest magnitude allows us to identify values of $\omega$ closest to $\omega_0$.

\subsection{Identifying solutions}

Our system will have $5N_\lambda N_\zeta$ eigenvalues, of which many result from the choice of discretisation and are not true eigenvalues of the continuous system. As such, there may be difficulties identifying the required solutions from the `numerical' ones. In practise, we find that this is rarely an issue, true solutions can often be found without requiring a good initial guess and can be distinguished from `numerical' ones by eye where needed. However, difficulties may arise when using large number of grid-points, an unstable background flow or a complicated topography. In this case, we use three main techniques for identifying solutions:
\begin{enumerate}
\item True eigenvalues may be identified by varying the number of grid-points as described in \cite{Johnson_Rodney_2011}. Eigenvalues of the continuous system will remain fixed while eigenvalues resulting from the discretisation are sensitive to changes in the grids.
\item A good initial guess for the frequency, $\Re[\omega]$, and growth rate, $\Im[\omega]$, allow for a good choice of $\omega_0$. The barotropic and equivalent-barotropic problems may be considered by replacing $z$ derivatives with $m\pi$ for some vertical wavenumber $m$. The resulting one-dimensional problem can be solved using the spectral collocation method on the offshore grid only. While we cannot study offshore variations in $H$ or depth-dependence in $U$, modes found using this simpler system may be used as an initial guess for the full system.
\item Parameter continuation can be used to follow a modal solution through parameter space. For example, we may start with a solution in the case of zero background flow and increase the magnitude of $U$, using the value of $\omega$ each time as the next initial guess. We often use this approach to determine $\omega$ as a function of varying wavenumber, $k$.
\end{enumerate}

Finally, we note that our approach---and indeed any numerical approach---is less effective at finding super-inertial waves with $|\omega|>|f|$ as the wave spectrum contains a continuous spectrum of Poincar\'e and inertial waves in this region. Therefore, the numerical results are dominated by modes with frequencies around $\omega \approx \pm f$. Since these waves do not decay as $y \to \infty$, they cannot be represented using exponentially-scaled Laguerre polynomials so often appear as unphysical, highly oscillatory solutions in our numerical spectrum. It is possible to identify CTW and instability modes in this region, provided an accurate initial guess for the frequency is known.

\section{Examples}
\label{sec:examples}

Here we consider three examples to demonstrate the effectiveness of our approach. The first two examples focus on idealised models of waves and instabilities whereas the third example shows how our spectral code can be used to find coastal-trapped wave modes in a realistic setup.

\subsection{Example 1: Kelvin wave-shelf wave transitions}

Here we consider a non-dimensional, idealised setup with $f = 1$, $N^2(z) = 1$, $\delta_h = 0$, $\delta_a = 0$ and
\begin{equation}
H(y) = H_0 + (1-H_0) \tanh y,
\end{equation}
where $H_0$ describes the fluid depth at the coast, $y = 0$. We use $N_\lambda$ points in the offshore direction, distributed as Laguerre points over the region $\lambda \in [0, 4]$. $N_\zeta$ Chebyshev points of the second kind are used in the stretched vertical grid. Note that in this non-dimensional setup, the alongshore wavenumber $k$ is nondimensionalised using the baroclinic Rossby radius, $L = NH/f$.

Taking $H_0 = 1$ allows us to compare our results with the theoretical result for Kelvin waves
\begin{equation}
\omega = \frac{k}{n\pi},
\end{equation}
where $n$ is the mode number, a positive integer corresponding to the number of zero crossings in the pressure field on the wall $y = 0$. Using just $(N_\lambda,N_\zeta) = (21,21)$ grid-points, we recover this result with a relative error of $10^{-8}$ for the first five modes. Larger grids are required to fully resolve the higher order modes due to their smaller spatial scales.

For $H_0 < 1$, we expect our solutions to consist of both Kelvin waves and shelf waves. As we will see, both of these type of wave correspond to the same solution curves, however they can have qualitatively different behaviour. Kelvin waves occur along vertical boundaries in the presence of a stratification. They are characterised by an exponential offshore decay and a phase speed, $c_p = \omega/k$, which is independent of $k$. By contrast, shelf waves occur due to sloping bottom topography. They are oscillatory in the offshore direction in regions of non-zero bottom slope, and have a phase speed which depends on $k$.

\begin{figure*}[h!]
\centering
\begin{subfigure}{0.45\textwidth}
\includegraphics[width=\textwidth,trim={0 0 0 0},clip]{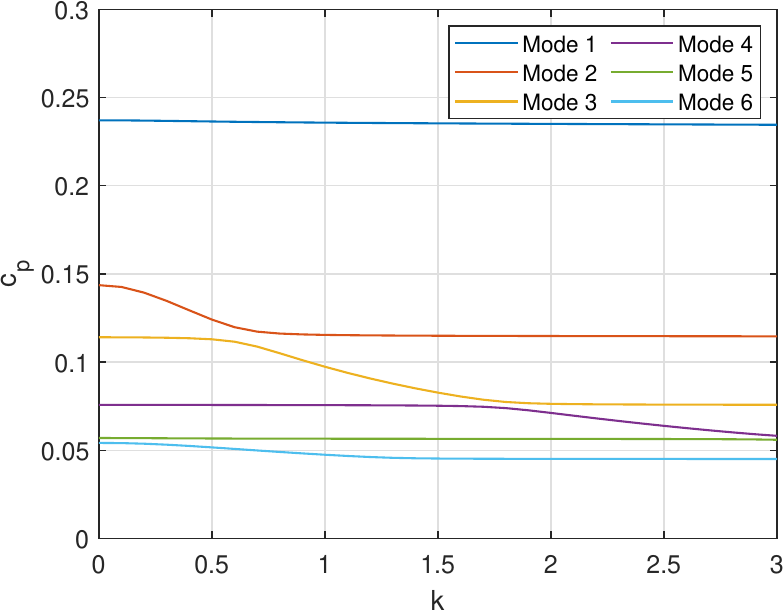}
\end{subfigure}
\begin{subfigure}{0.45\textwidth}
\includegraphics[width=\textwidth,trim={0 0 0 0},clip]{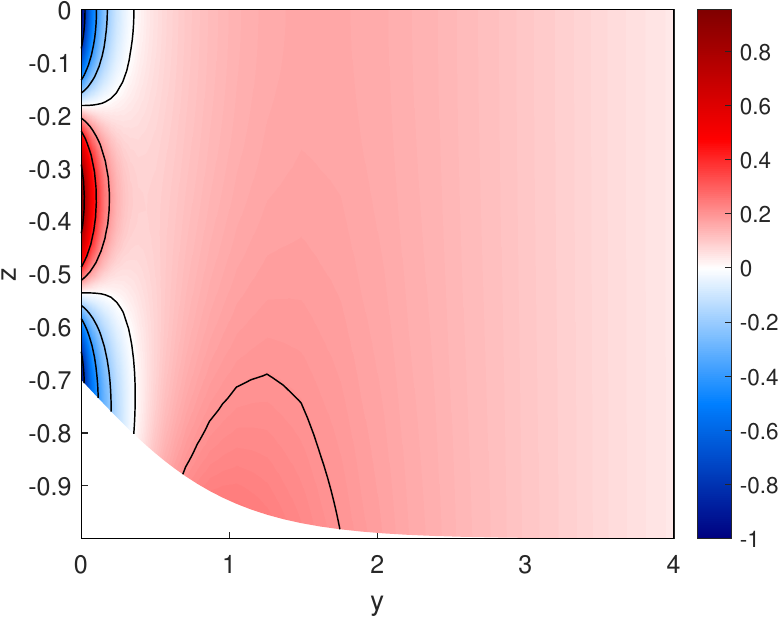}
\end{subfigure}
\\
\begin{subfigure}{0.45\textwidth}
\includegraphics[width=\textwidth,trim={0 0 0 0},clip]{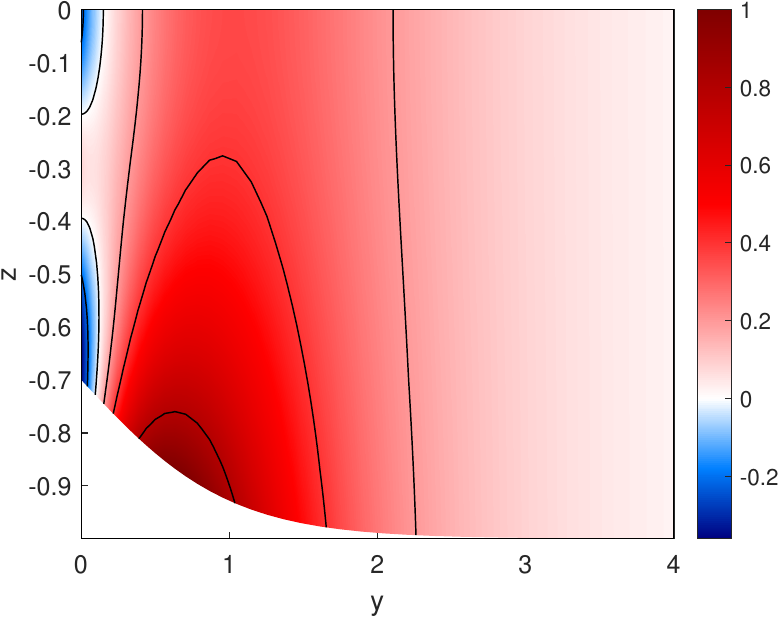}
\end{subfigure}
\begin{subfigure}{0.45\textwidth}
\includegraphics[width=\textwidth,trim={0 0 0 0},clip]{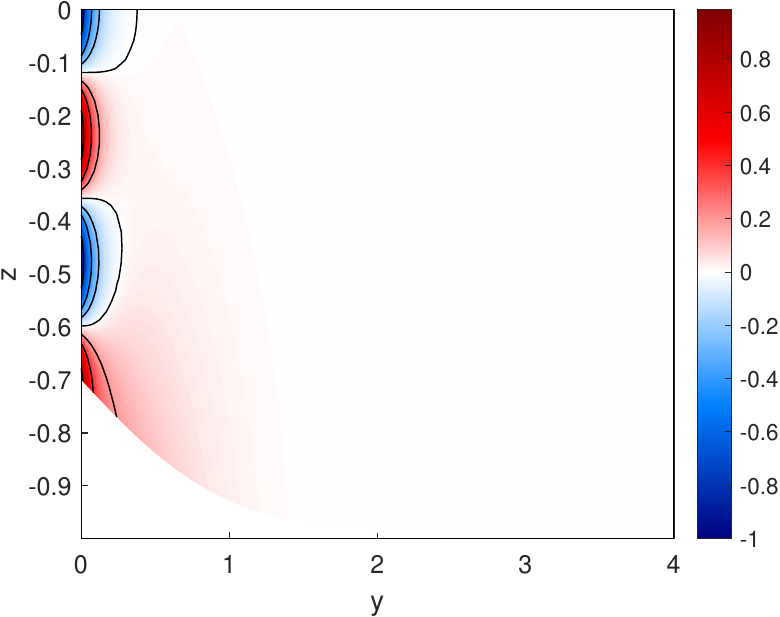}
\end{subfigure}
\caption{Numerical CTW modes for $H_0 = 0.7$, $f = 1$, $N^2 = 1$ and $\delta_h = \delta_a = 0$. (a) Plots of the phase speeds, $c_p$, of the first 6 modes as a function of alongshore wavenumber, $k$. (b)-(d) Plots of the pressure eigenfunctions, $\hat{p}(y,z)$, as functions of $y$ and $z$ for the mode $3$ solution with $k = 0.1$ (b), $k = 1$ (c) and $k = 3$ (d).}
\label{fig:mod3}
\end{figure*}

\cref{fig:mod3} shows numerical results for $H_0 = 0.7$. Panel (a) shows the phase speeds for the first $6$ modes as a function of the alongshore wavenumber $k$. We observe regions where $c_p$ is approximately independent of $k$ (Kelvin-wave-like) and regions where $c_p$ decreases with $k$ (shelf-wave-like);  transitions between these regions occur at near-intersections between neighbouring curves. These near-intersections correspond to transitions between shelf wave and Kelvin wave behaviour and have been previously studied by \cite{Allen_75,Huthnance_78}. Rather than the phase speed curves crossing, the two modes exchange `identities', with Kelvin waves becoming shelf waves and vice versa. This can be best seen in the mode $3$ curve where an initially flat $c_p$ curve transitions to a $k$ dependent curve around $k = 0.7$ before returning to $k$ independence around $k = 1.8$. Panels (b)-(d) show the pressure eigenfunction $\hat{p}(y,z)$ for the mode 3 solution at wavenumbers $k = 0.1$ (a), $k = 1$ (b), and $k = 3$ (d). We can clearly see that panels (b) and (d) resemble Kelvin waves trapped on the wall $y = 0$ whereas (c) has a strong pressure signal on the bottom boundary characteristic of a shelf wave. Note however that the number of pressure extrema remains unchanged so all cases correspond to the same mode. Further, we observe that the boundary acts to `tilt' the modes such that the contour of zero pressure intersects with the surface. This observation is consistent with previous studies \cite{Huthnance_78,Crowe_Johnson_2020}.

Asymptotically, this behaviour may be understood as follows. For small $k$, the offshore decay scale is large so the shelf region is close to a vertical wall on this scale. Therefore, the wave behaves as a Kelvin wave over the full depth, $z \in [-1,0]$, with phase speed $c_p = 1/(n\pi)$. Conversely, for large $k$, the offshore decay scale is large so the slope acts as a flat horizontal boundary at $z = -H_0$. Therefore, the wave behaves as a Kelvin wave on the region $z\in [-H_0,0]$ with the reduced phase speed $c_p = H_0/(n\pi)$. For intermediate $k$, the offshore decay scale can match the horizontal scale of the shelf region, leading to a regime where the wave becomes a shelf wave. Here $c_p$ depends strongly on $k$ as this parameter determines the ratio between the offshore scale of the wave and the topographic length-scale. For higher order modes, with multiple pressure extrema on the shelf region, we expect a complicated series of transitions as each pressure extrema moves from the shelf region up onto the wall as $k$ increases. This corresponds to transitions between different shelf-wave modes.

\begin{figure*}[h!]
\centering
\begin{subfigure}{0.45\textwidth}
\includegraphics[width=\textwidth,trim={0 0 0 0},clip]{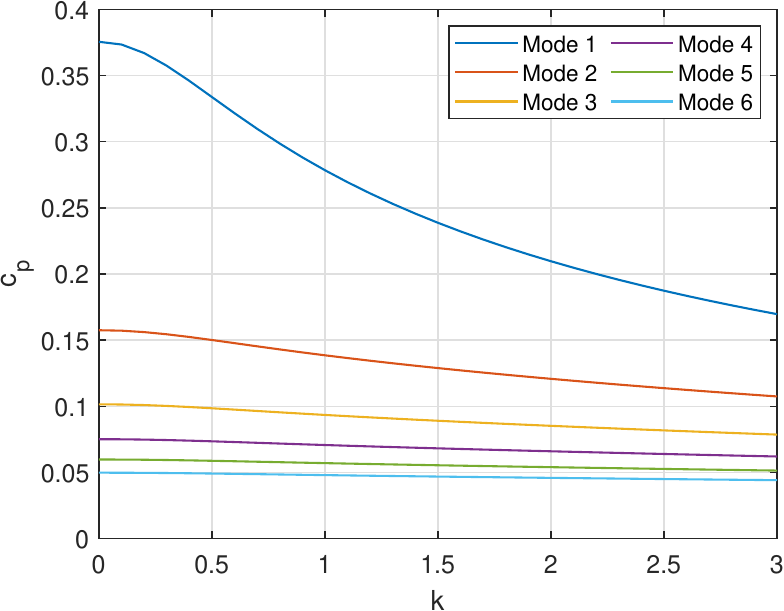}
\end{subfigure}
\begin{subfigure}{0.45\textwidth}
\includegraphics[width=\textwidth,trim={0 0 0 0},clip]{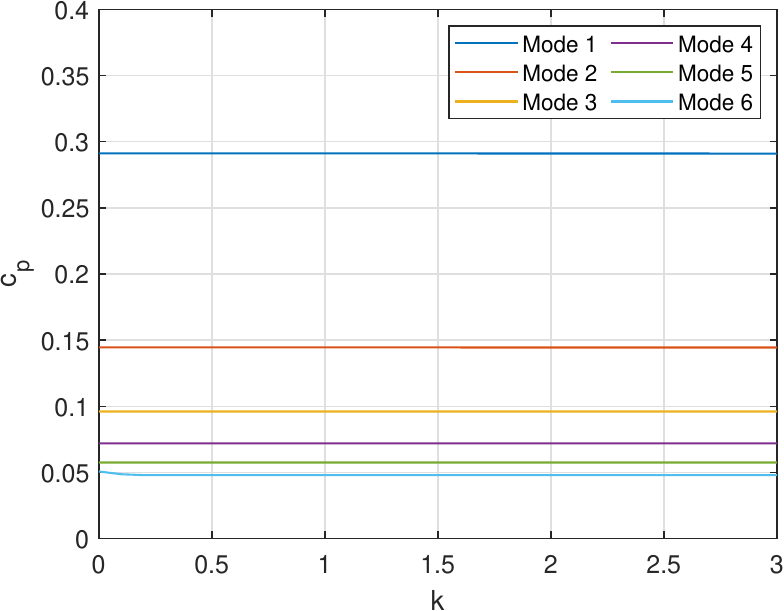}
\end{subfigure}
\\
\begin{subfigure}{0.45\textwidth}
\includegraphics[width=\textwidth,trim={0 0 0 0},clip]{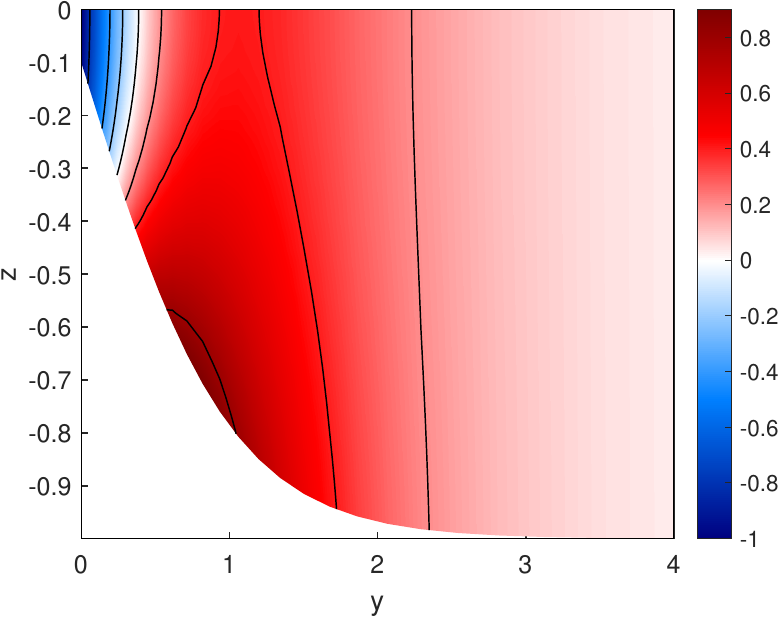}
\end{subfigure}
\begin{subfigure}{0.45\textwidth}
\includegraphics[width=\textwidth,trim={0 0 0 0},clip]{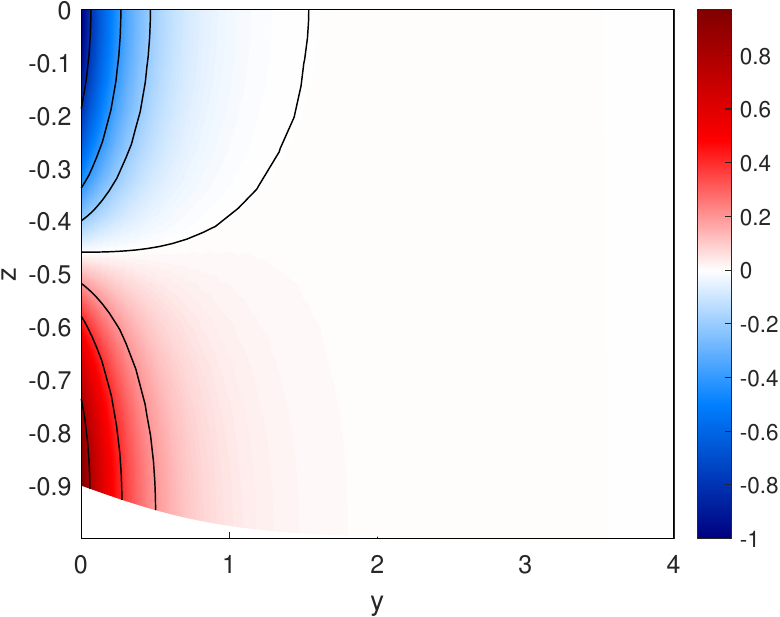}
\end{subfigure}
\caption{Panels (a) and (b): plots of the phase speed, $c_p$, for the first 6 modes as a function of alongshore wavenumber, $k$, for coastal depths (a) $H_0 = 0.1$ and (b) $H_0 = 0.9$. Panels (c) and (d): plots of the pressure eigenfunction, $\hat{p}(y,z)$, as functions of $y$ and $z$ with $k = 1$ for (c) $H_0 = 0.1$ and (d) $H_0 = 0.9$. All solutions are shown for $f = 1$, $N^2 = 1$ and $\delta_h = \delta_a = 0$.}
\label{fig:H0_comp}
\end{figure*}

\cref{fig:H0_comp} (a) and (b) show the phase speed as a function of $k$ for $H_0 = 0.1$ and $H_0 = 0.9$. For $H_0 = 0.1$ we observe that all solutions behave as shelf waves due to the increased size of the sloping region. By contrast, for $H_0 = 0.9$, the solutions behave predominantly as Kelvin waves with phase speeds given approximately by $c_p = H_0/(n\pi)$. Panels (c) and (d) show the pressure eigenfunction, $\hat{p}(y,z)$, for both values of $H_0$, allowing the distinction between shelf wave and Kelvin wave structure to be seen.

\subsection{Example 2: Quasi-barotropic instability of a coastal shear flow}

\begin{figure*}[h!]
\centering
\begin{subfigure}{0.45\textwidth}
\includegraphics[width=\textwidth,trim={0 0 0 0},clip]{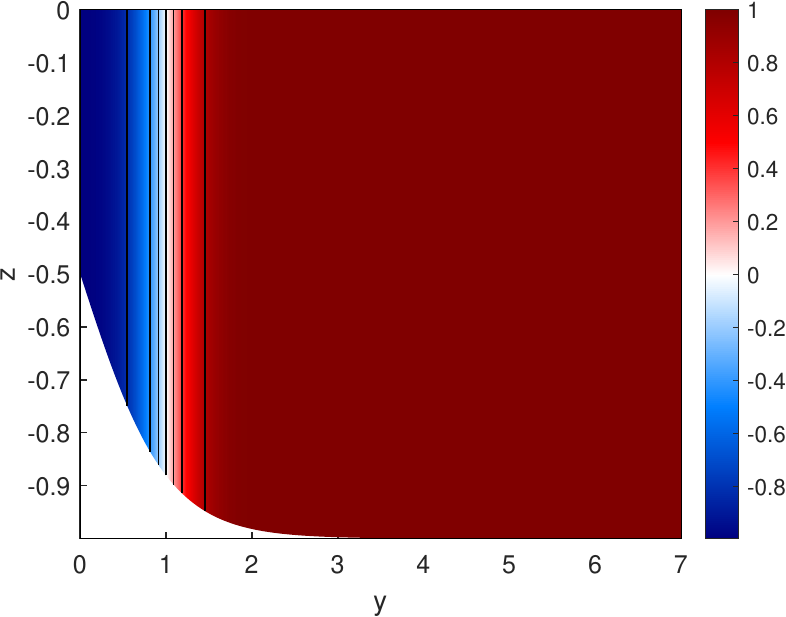}
\end{subfigure}
\begin{subfigure}{0.45\textwidth}
\includegraphics[width=\textwidth,trim={0 0 0 0},clip]{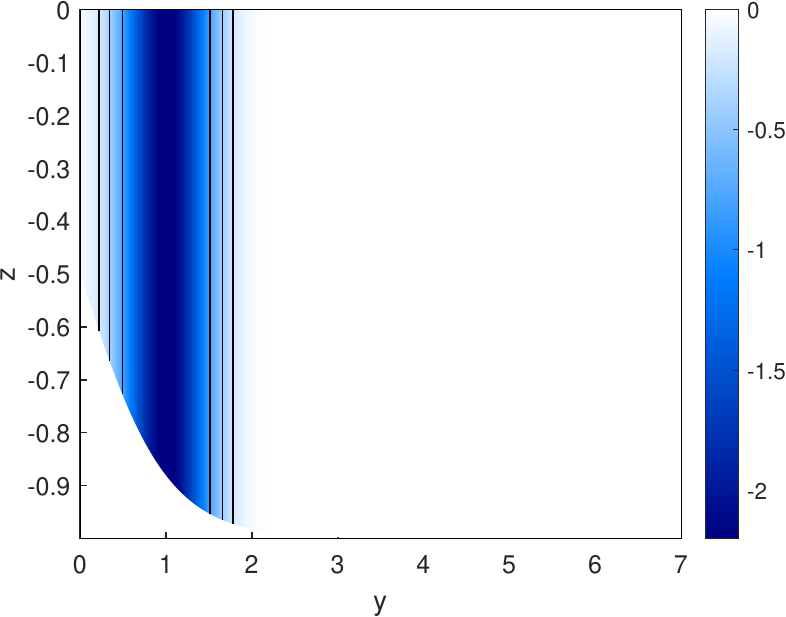}
\end{subfigure}
\caption{The background velocity $U(y,z)$, panel (a), and background vorticity $-U_y(y,z)$, panel (b), for $H_0 = 0.5$.}
\label{fig:U_plot}
\end{figure*}

In order to consider an unstable setup we introduce a background flow
\begin{equation}
U(y,z) = \erf[2(y-1)],
\end{equation}
and use the same parameters as the previous example; $f = 1$, $N^2(z) = 1$, $\delta_h,\delta_a = 0$ and 
\begin{equation}
H(y) = H_0 + (1-H_0)\tanh y.
\end{equation}
Here, $U$ is taken to be depth-independent for simplicity and corresponds to a horizontally-sheared, along coast flow with a jet in the vertical vorticity centred at $y = 1$. Non-trivial vertical structure will enter the instability mode solutions due to the sloping bottom topography. In order to fully resolve the instability within the offshore vorticity jet, we take a composite grid in the $y$ direction consisting of a segment of Chebyshev points for $y \in [0,1]$ and a segment of Laguerre points for $y \in [1,7]$. The background flow, $U$, and vorticity, $-U_y$, for $H_0 = 0.5$ are shown in \cref{fig:U_plot} as functions of $y$ and $z$. We note that a constant background flow $U_0$ can be added to $U$ to set the value of $U$ on the coast or in the far field. This addition will Doppler shift the frequency, $\omega \to \omega + U_0 k$, but will not affect the structure or growth rate of any solutions.

To demonstrate the effectiveness of our method for finding instability modes we will look for `quasi-barotropic' instability modes to this problem. We define `quasi-barotropic' modes to be solutions to our system which approach barotropic instability modes in the constant depth ($H_0 = 1$) case. These modes have a strong pressure signal on the wall $y = 0$ which is used to differentiate them from the various baroclinic solutions also admitted by the system. The barotropic instability solutions are found with a 1D barotropic code which solves the problem from \cref{sec:num_meth} under the assumption of no $z$ dependence (and hence no vertical velocity or buoyancy perturbations). Once the barotropic modes are found, we can use a parameter continuation approach by reducing $H_0$ and using the value of $\omega$ for the previous $H_0$ as our initial guess $\omega_0$.

\begin{figure*}[h!]
\centering
\begin{subfigure}{0.45\textwidth}
\includegraphics[width=\textwidth,trim={0 0 0 0},clip]{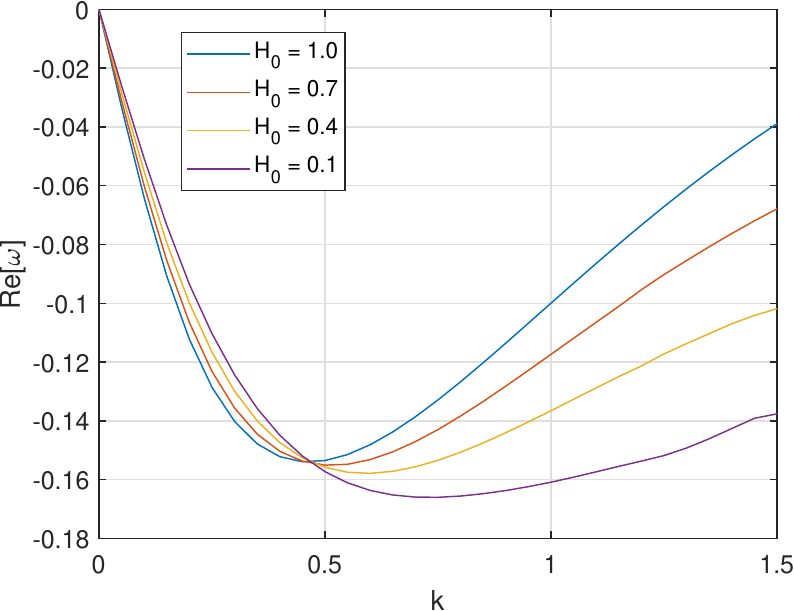}
\end{subfigure}
\begin{subfigure}{0.45\textwidth}
\includegraphics[width=\textwidth,trim={0 0 0 0},clip]{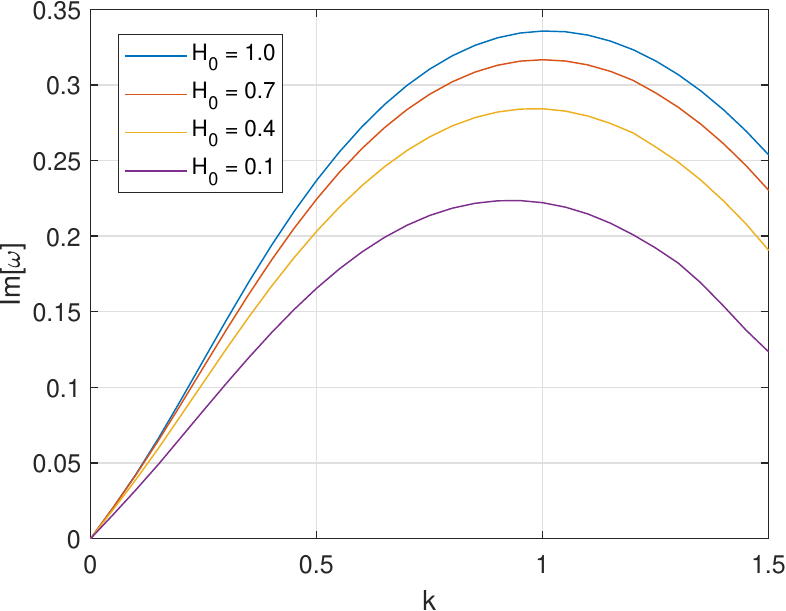}
\end{subfigure}
\caption{The real (a) and imaginary (b) parts of $\omega$ as a function of alongshore wavenumber $k$ for four different values of coastal depth $H_0$. The real part of $\omega$ describes the frequency while the imaginary part is the instability growth rate.}
\label{fig:omega_k_instab}
\end{figure*}

\cref{fig:omega_k_instab} shows the frequency, $\Re[\omega]$, and growth rate, $\Im[\omega]$, as a function of alongshore wavenumber, $k$, for $H_0 \in \{1.0,0.7,0.4,0.1\}$. We can see that the fastest growing mode occurs for $k \approx 1$ and this $k$ value reduces slightly as $H_0$ decreases. Further, the maximum growth rate is reduced by around 40\% between $H_0 = 1$ and $H_0 = 0.1$ with this growth rate reduction corresponding to an increase in the frequency, and hence phase speed, of the mode. This is consistent with the results of \cite{Kuo_1949} where a sufficiently large background potential vorticity gradient has a stabilising effect on the modes.

\begin{figure*}[h!]
\centering
\begin{subfigure}{0.45\textwidth}
\includegraphics[width=\textwidth,trim={0 0 0 0},clip]{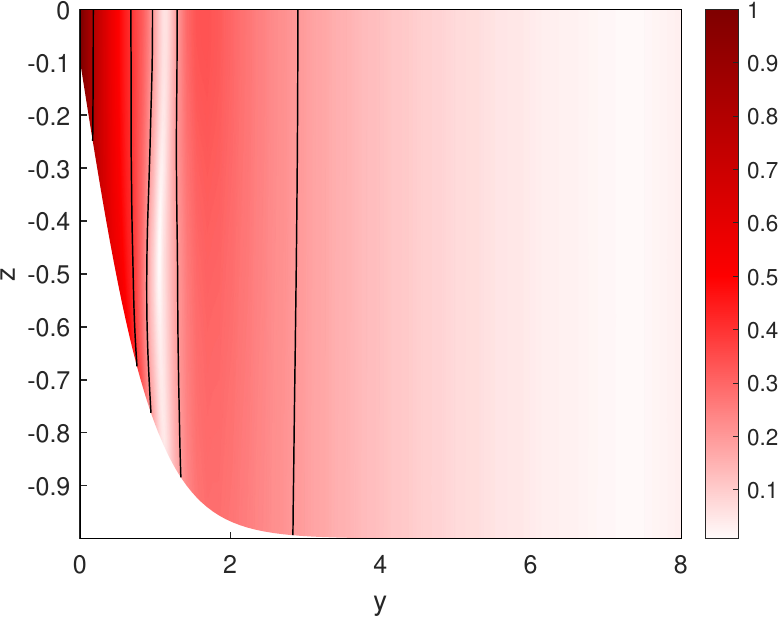}
\end{subfigure}
\begin{subfigure}{0.45\textwidth}
\includegraphics[width=\textwidth,trim={0 0 0 0},clip]{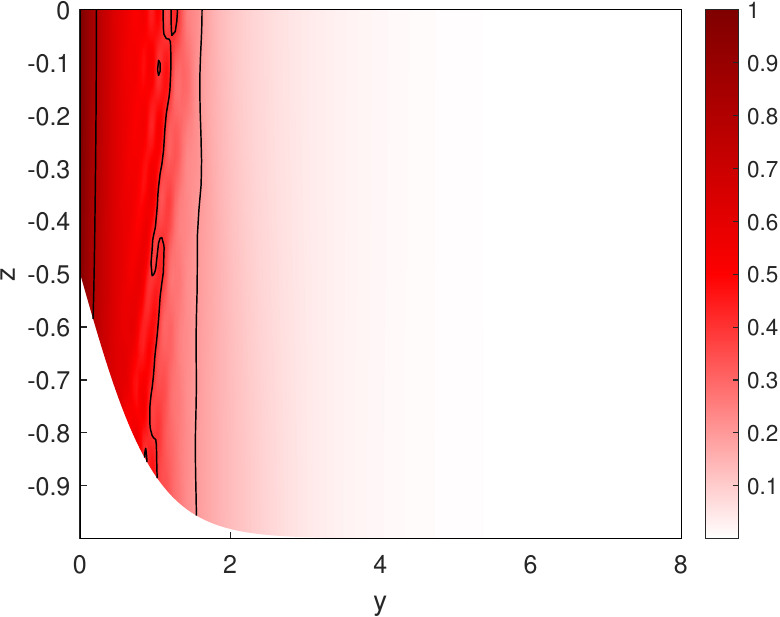}
\end{subfigure}
\caption{The magnitude of the pressure eigenfunction, $|\hat{p}(y,z)|$, as a function of $y$ and $z$ for (a) $(k,H_0) = (0.4,0.1)$ and (b) $(k,H_0) = (1,0.5)$. Solutions are shown for $f = 1$, $N^2 = 1$ and $\delta_h = \delta_a = 0$.}
\label{fig:instab_abs_p}
\end{figure*}

\cref{fig:instab_abs_p} shows the magnitude of the pressure eigenfunction, $|\hat{p}|$, as a function of $y$ and $z$. Small vertical variation is induced by the condition of no flow through the tilted boundary. We observe that increasing the slope gradient (by decreasing $H_0$) does not necessarily correspond to an increase in the small-scale vertical variation in the modal solution. Rather, vertical variation appears to be induced if the value of $\omega$ of the quasi-barotropic mode is close to that of a mode in the baroclinic spectrum. In this case, the quasi-barotropic mode appears to contain small scale structure similar to that of the corresponding baroclinic mode.

\subsection{Example 3: Finding realistic CTW modes}
\label{sec:ex3}

Gelderloos et. al. \cite{Gelderloos_et_al_21} used realistic numerical simulations to examine subinertial variability along the Southeast Greenland coast. They found strong CTW signals within two frequency bands which were theorised to correspond to the first and third modes in the CTW spectrum (referred to as mode-I and mode-III respectively). The mode-I signal was shown to correspond to the first CTW mode using the approach of Brink \cite{Brink_1982,Brink_2006} however this method was unable to identify the mode-III CTW solution. Here, we consider the setup of \cite{Gelderloos_et_al_21} and show that our method is capable of identifying this mode-III solution. Our results are consistent with the frequency bands obtained in \cite{Gelderloos_et_al_21} and have a spatial structure which is consistent with the EOF analysis there.

\begin{figure*}[h!]
\centering
\begin{subfigure}{0.45\textwidth}
\includegraphics[width=\textwidth,trim={0 0 0 0},clip]{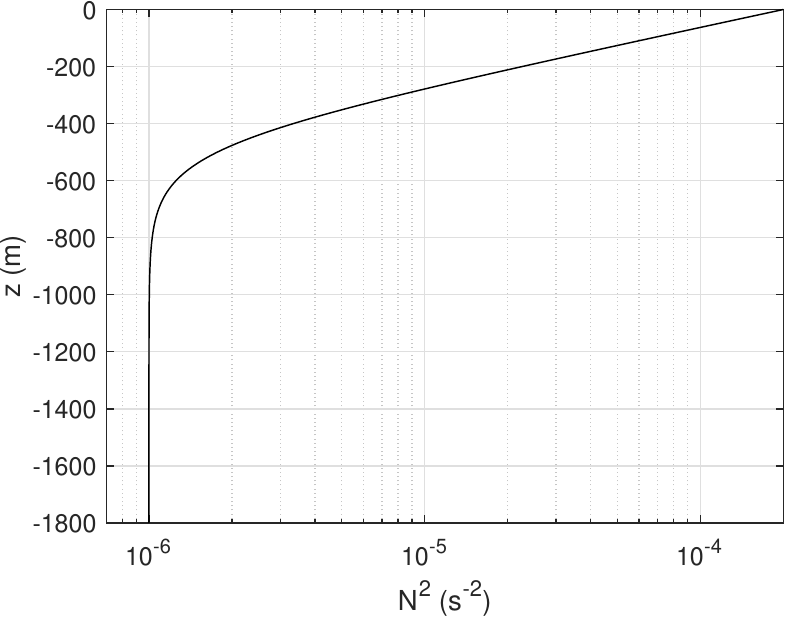}
\end{subfigure}
\begin{subfigure}{0.45\textwidth}
\includegraphics[width=\textwidth,trim={0 0 0 0},clip]{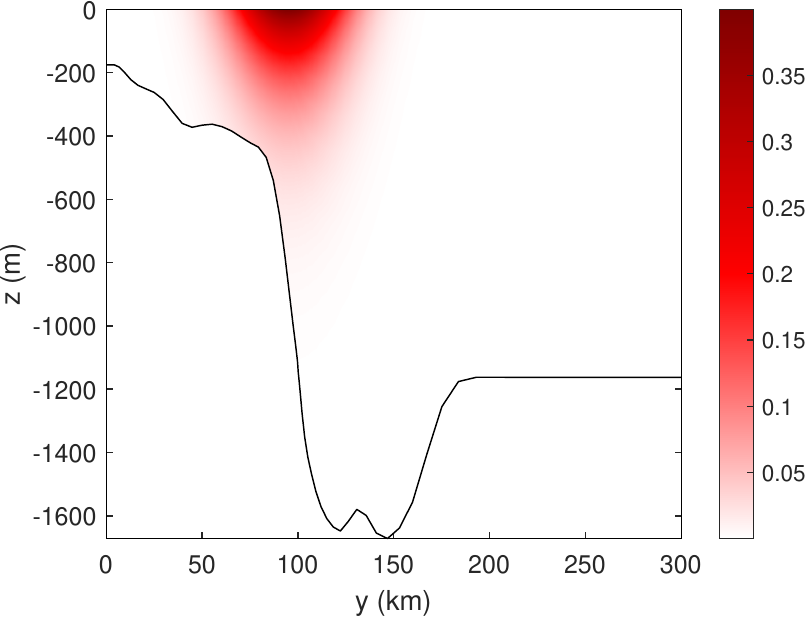}
\end{subfigure}
\caption{Background state corresponding to the setup of \cite{Gelderloos_et_al_21}. (a) The buoyancy stratification, $N^2$, as a function of $z$. (b) The along-coast background velocity field, $U(y,z)$, as a function of $y$ and $z$ (color)  and the coastal topography, $H(y)$ (black line).}
\label{fig:Ex_3_Setup}
\end{figure*}

We take a shelf profile, $H = H(y)$, which corresponds to the cyan section in Fig. 1 of \cite{Gelderloos_et_al_21}. A 5-point moving average filter is applied to prevent grid-scale oscillations in our numerical results and the profile is matched to a constant depth region for large $y$ to ensure that the mode decays away from the coastline. We take an offshore grid consisting of Chebyshev points on the region $y \in [0\unit{km},100\unit{km}]$ and Laguerre points in the region $y \in [100\unit{km},300\unit{km}]$. A single Chebyshev grid is taken in the vertical ($z$) direction. We take a background flow of
\begin{equation}
U(y,z) = U_0\exp(-(y-L_c)^2/L_w^2)\exp(z/L_z),
\end{equation}
and stratification
\begin{equation}
N^2(z) = N_s^2\exp(z/L_N)+N_d^2,
\end{equation}
which correspond to exponential fits to the velocity and summer stratification identified by \cite{Gelderloos_et_al_21} with parameters $U_0 = 0.4\unit{m/s}$, $L_c = 95\unit{km}$, $L_w = 30\unit{km}$, $L_z = 200\unit{m}$, $N_s^2 = 2\times 10^{-4}\unit{s$^{-2}$}$, $L_N = 90\unit{m}$, and $N_d^2 = 10^{-6}\unit{s$^{-2}$} $. \cref{fig:Ex_3_Setup} shows the buoyancy stratification $N^2$, panel (a), and the topography $H$ and background velocity $U$, panel (b). We consider non-hydrostatic solutions with a free surface, $(\delta_h, \delta_a) = (1,1)$, and use a Coriolis frequency of $f = 1\times 10^{-4}\unit{m/s}^2$ and a gravitational acceleration of $g = 10\unit{m/s}^2$.

\begin{figure*}[h!]
\centering
\begin{subfigure}{0.45\textwidth}
\includegraphics[width=\textwidth,trim={0 0 0 0},clip]{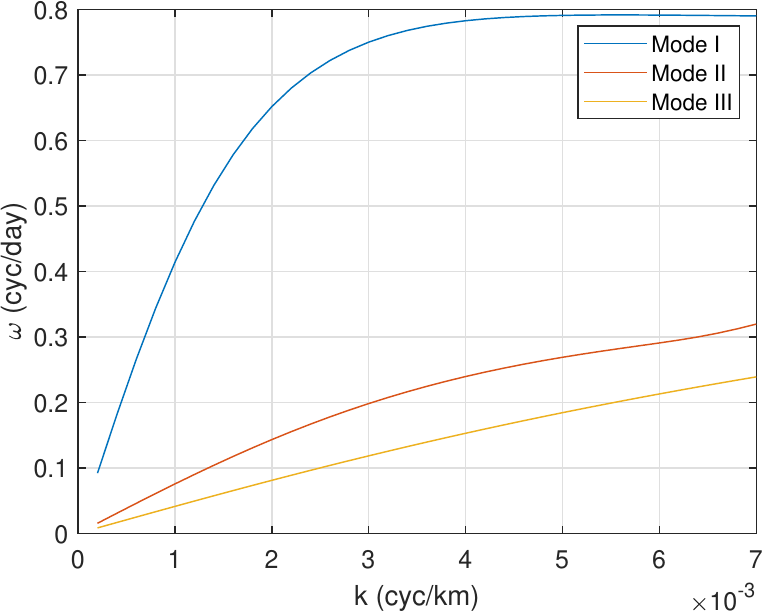}
\end{subfigure}
\begin{subfigure}{0.45\textwidth}
\includegraphics[width=\textwidth,trim={0 0 0 0},clip]{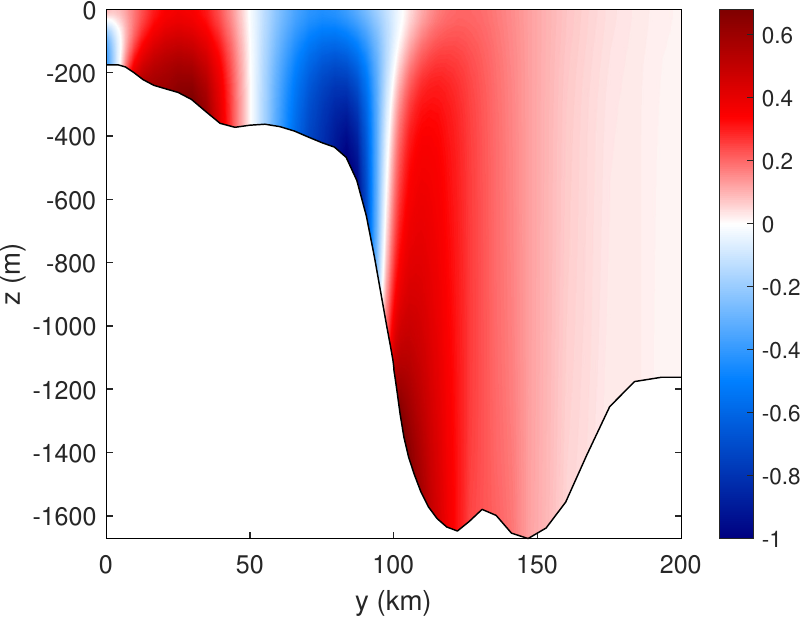}
\end{subfigure}
\caption{Solutions for the CTW solutions over realistic topography. (a) The frequency, $\omega$, as a function of alongshore wavenumber for the first three modes. (b) The pressure eigenfunction, $\hat{p}(y,z)$, as a function of $y$ and $z$ for the mode-III solution at $k = 5\times 10^{-3} \unit{cyc/km}$. These solutions use $f = 1\times 1^{-4} \unit{s}^{-1}$, $\delta_h = 1, \delta_a = 1$ and the background state of \cref{fig:Ex_3_Setup}.}
\label{fig:Ex_3_omega_modes}
\end{figure*}

\cref{fig:Ex_3_omega_modes}.(a) shows the frequency, $\omega$, of the first three CTW modes for alongshore wave-numbers up to $k  = 7\times 10^{-3} \unit{cyc/km}$. The mode-I and mode-III curves are consistent with the wavenumber and frequency of the two subinertial bands identified in  \cite{Gelderloos_et_al_21}. These solution curves are obtained by examining the CTW solutions for $k  = 2\times 10^{-4} \unit{cyc/km}$ and following the desired modes as $k$ is increased, using a second-order central difference method to estimate the value of $\omega_0$ from previous $\omega$ values. We remove unphysical solutions by selecting only modes with a purely real frequency where the integrated difference in the pressure eigenfunction does not exceed a specified tolerance.

\cref{fig:Ex_3_omega_modes}.(b) shows the pressure eigenfunction, $\hat{p}(y,z)$, for the mode-III solution with $k = 5\times 10^{-3} \unit{cyc/km}$, corresponding to the predicted mode-III wave in \cite{Gelderloos_et_al_21}. We note that the frequency of this solution is within the error bars given by \cite{Gelderloos_et_al_21} while the surface pressure anomaly is consistent with the mode-III EOF identified from an analysis of sea-surface height and surface velocity in the region of interest. As noted by \cite{Gelderloos_et_al_21}, our results depend only weakly on the background flow profile and stratification, with the speed and structure of the modes being set primarily by the Coriolis frequency and shelf geometry. Further, our results are found to be insensitive to the smoothing of the shelf profile, $H$, provided the grid resolution is sufficient to resolve any ridges or seamounts.

\section{Discussion and Conclusions}
\label{sec:diss_conc}

Here we have presented a method for finding CTW modes and instabilities over coastal topography in the presence of a background flow. This work extends the method of \cite{Johnson_Rodney_2011} to include non-hydrostatic effects, an alongshore background flow and an offshore grid which can be customised to suit the shelf geometry.

Our method is shown to effectively replicate previous theoretical and numerical results across a range of problems. Here, we have shown three examples, covering idealised shelf and Kelvin waves, idealised instabilities of a coastal shear flow and CTW solutions with realistic ocean parameters and a complicated coastal topography. In all cases, the required modes can be identified by applying our method to a low resolution grid and examining the set of solutions. Higher resolution spatial structure can be obtained by using the low resolution frequency as an initial guess for a higher resolution run while frequency, or phase speed, curves can be obtained by increasing the wavenumber and estimating the next frequency using previous points on the curve.

As wave and instability modes are identified as eigenvectors of a discretised linear system, a large set of modes can be quickly identified using standard methods for solving generalised eigenvalue problems. The main disadvantage of a discrete formulation of this system is the occurrence of `numerical' eigenvalues, i.e. those eigenvalues which correspond to the discretisation rather than the underlying continuous problem. The use of spectral methods for calculating derivatives enables us to take advantage of the exponential convergence of such methods and hence reduce the number of grid-points required. This low grid resolution reduces the number of numerical eigenvalues and makes identifying `true' solutions easier. Additional methods for identifying `true' solutions include varying the grid resolution to determine grid-independent modes \cite{Johnson_Rodney_2011} and limiting the area-integrated change in spatial structure when tracing solutions through wave-number space.

This method can be applied to a wide range of linear wave/instability problems across geophysical fluid dynamics and applied mathematics. The Matlab implementation of this method is included as supplementary material, along with example scripts demonstrating the three examples considered in \cref{sec:examples}.

\section*{Acknowledgements}

The authors would like to thank Dr. R. Gelderloos for helpful discussions of the coastal-trapped waves observed in the simulations of \cite{Gelderloos_et_al_21} and for providing the shelf topography profile used in \cref{sec:ex3}.




\bibliographystyle{elsarticle-num} 
\bibliography{cas-refs}

\end{document}